\documentclass[aps,twocolumn,floatfix,prb,a4paper,longbibliography]{revtex4-2}

\usepackage{natbib}
\usepackage[%
  colorlinks=true,
  urlcolor=blue,
  linkcolor=red,
  citecolor=blue
]{hyperref}
\usepackage{color}
\usepackage{graphicx}
\usepackage{amsmath}
\usepackage{amsfonts}
\usepackage{bm}

\begin{document}

\title{Efficient impurity-bath trial states from superposed Slater determinants}
\author{Izak Snyman}
\affiliation{Mandelstam Institute for Theoretical Physics, School of Physics, University of the 
Witwatersrand, Johannesburg, South Africa}
\author{Serge Florens}
\affiliation{Institut N\'{e}el, CNRS and Universit\'e Grenoble Alpes, F-38042 Grenoble, France}

\begin{abstract}
The representation of ground states of fermionic quantum impurity problems as superpositions
of Gaussian states has recently been given a rigorous mathematical foundation. 
[S. Bravyi and D. Gosset, Comm. Math. Phys. 356, 451 (2017)]. It is natural to ask how many
parameters are required for an efficient variational scheme based on this representation.
An upper bound is $\mathcal O(N^2)$, where $N$ is the system size, which corresponds to
the number parameters needed to specify an arbitrary Gaussian state.
We provide an alternative representation, with more favorable scaling, only requiring
$\mathcal O(N)$ parameters, that we illustrate for the interacting resonant level model.
We achieve the reduction by associating mean-field-like parent Hamiltonians with the individual 
terms in the superposition, using physical insight to retain only the most relevant channels 
in each parent Hamiltonian.
We benchmark our variational ansatz against the Numerical Renormalization Group, and compare our 
results to existing variational schemes of a similar nature to ours. Apart from the ground state energy, 
we also study the spectrum of the correlation matrix -- a very stringent measure of accuracy.
Our approach outperforms some existing schemes and remains quantitatively accurate in the numerically 
challenging near-critical regime.
\end{abstract}

\maketitle

\section{Introduction}
\subsection{Motivation}
The brute force diagonalization of a generic quantum many-body problem 
requires computational 
resources that grow exponentially with the system size, and is therefore impracticable for 
systems with more than a handful of particles.
A major achievement in the field of strongly correlated electrons has been the development of
numerical methods for special classes of problems that circumvent this exponential barrier.
Examples include the Numerical Renormalization Group (NRG)~\cite{Bulla2008} 
for quantum impurity problems, 
the Density Matrix Renormalization Group~\cite{White_DMRG,Schollwock_Review} for
one-dimensional lattice problems, and Quantum Monte Carlo 
simulations~\cite{Gull} in situations where the ``sign problem'' is managable.
However, these 
numerical methods typically do not give direct access to useful expressions for correlated ground states 
in terms of the bare degrees of freedom appearing in the microscopic Hamiltonian. 
Instead, the ground state structure must be inferred from the sometimes limited set of
observables the method allows one to calculate.

It is therefore desirable to develop methods that provide intuition for the nature of correlated
ground states, even in cases where existing methods provide a numerically 
exact answers~\cite{Barcza}. A well-established approach in this context revolves around 
parent Hamiltonians~\cite{Greiter}. The idea is to identify a Hamiltonian whose ground state can
be computed easily and can serve as an idealization of the correlated state of interest. 
Since operators do not have to be ``close'' to each other to have similar behavior in a restricted 
subspace, the parent Hamiltonian may be very different from the microscopic one, thus providing
a new and useful perspective on the ground state in question.

An almost trivial example of the parent Hamiltonian idea is Hartree-Fock mean field theory, in
which an interacting Hamiltonian is replaced by an optimal non-interacting approximation. 
Mean field theory often serves to identify the types of behavior that a system may
host.
However, there are well-known examples where mean field theory severely overestimates the ground
state energy, or predicts spontaneous symmetry breaking when the true ground state is symmetric.
This is typically the case for quantum impurities where even local perturbations can hybridize 
distinct symmetry broken states.
Furthermore, mean-field parent Hamiltonians cannot produce the non-Gaussian
correlation functions that often characterize interacting problems. 
Going beyond non-interacting parent Hamiltonians requires ingenuity 
and a case-by-case approach. This explains why important classes of many-body problems
have to date not benefitted from the parent Hamiltonian method.
 
In this article we focus on one such class, namely quantum impurity models. 
In this context, a favored method involves
constructing variational trial states by forming linear combinations of Gaussian states.
Approximations with linear combinations of two Gaussians were already used long ago 
to obtain a qualitative physical description~\cite{Anderson, Emery,Silbey}.
Adding more terms in the superposition offers a viable route to arbitrary accuracy, as was first 
demonstrated systematically for the spin-boson model~\cite{Bera1,Bera2} and the scaling limit of
the anisotropic Kondo Hamiltonian~\cite{Snyman}.
Recently, it was proven analytically that the ground state of an arbitrary fermionic quantum
impurity problem can be approximated as a superposition of $M$ Gaussian states,
using computational resources that scale quasi-polynomially in the inverse of the 
accuracy~\cite{Bravyi}. 
Practical algorithms involve minimizing the expectation value of the energy over the
weights of the terms in the superposition and over the parameters of the Gaussian states.  Recently,
an efficient optimization method has been used~\cite{Bauer} to investigate spatial correlations in
the notoriously difficult two-channel Kondo problem. This and other existing methods
consider each Gaussian state to be completely arbitrary, which means that the number of
variational parameters per Gaussian state scales quadratically with the system size. 
It seems likely that physical insight into the specific system under consideration
could be exploited to reduce the number of parameters per Gaussian state.
Given the complexity
of finding the absolute minimum in a large scale non-convex optimization problem, 
such a reduction, if possible, could prove invaluable.
However, until now this has not been investigated.

\subsection{Superposed Gaussians from restricted parent Hamiltonians}
In this article, we demonstrate a strategy to significantly reduce the number of variational parameters 
per state in Gaussian superpositions.
We do so by parametrizing the Gaussian states using non-interacting parent Hamiltonians. 
This makes the physical meaning of the parameters transparent, and allows us to use
significantly fewer variational parameters than are required to specify an arbitrary Gaussian state.
We demonstrate our ideas by studying the Interacting Resonant Level Model (IRLM)~\cite{Vigman1978}, the simplest quantum impurity problem
in which an electronic impurity orbital interacts with a conduction band to produce Kondo 
correlations. 
In the past, the IRLM has provided an interesting test bed for studying 
equilibrium~\cite{Boulat2008a,Borda2008,Braun2014,Camacho2019} and 
dynamical~\cite{Doyon2007,Boulat2008b,Borda2010,Karrasch2010,Andergassen2011,Kennes2013,Freton2014,Anders2014,Weichselbaum2018,Arrigoni2019,Bidzhiev2019} 
features of impurity models.
Our key insight is the following.
When an electronic impurity interacts with a conduction band, there are two obvious channels involved.
The first is the Hartree channel, where the charge or spin density on the impurity induces
a site-dependent inhomogenous charge or spin density in the conduction band.
The second is the Fock channel, which leads to a hybridization of the impurity
and the conduction band orbitals. We take only these two channels into account when constructing
parent Hamiltonians.
As a result, the number of variational parameters in our approach only scales linearly with the system size,
as opposed to quadratically. We nonetheless  
obtain results that are significantly more accurate than some competing variational approaches.
We also show that pairing-type correlations~\cite{AshidaA,AshidaB}, do not play a major role for the model under consideration, so that the variational state can be constructed from Slater determinants, rather than more general Gaussians that host BCS correlations.

In addition, we emphasize here general symmetry considerations. 
This provides an alternative perspective on the use of superpositions
of a few well-chosen Slater determinants as variational trial states.
An important ingredient of quantum impurity models is 
particle-hole symmetry. While it is only obeyed for a fine-tuned electrostatic potential on the impurity, 
particle-hole symmetry breaking terms become irrelevant at Kondo correlated fixed points.
If we attempt to approximate the ground
state of a fermionic quantum impurity problem by a single Slater determinant $\left|F\right>$, the
generic structure of $\left|F\right>$ will be a Fermi sea in which plane wave orbitals are replaced
by scattering states in the presence of a static impurity. Very often, $\left|F\right>$ will break a
symmetry that the interacting Hamiltonian $H$ preserves. For simplicity, let us consider a
$\mathbb{Z}_2$
symmetry such as the above-mentioned particle-hole conjugation, and take its unitary and hermitian generator to be $P$.
Then $PHP=H$, but $P\left|F\right>\not=\left|F\right>$. If the true ground state preserves the
symmetry, one may conjecture a ground state approximation of the form
$\left|\psi\right>=\left|F\right>\pm P\left|F\right>$, which we call a ``Slater pair'', where the
two states in the superposition are related to each other by particle-hole conjugation.
Note here that the single particle orbitals
that are naturally associated with $P\left|F\right>$ are different from, but not orthogonal to
those associated with $\left|F\right>$. It therefore typically requires an enormous number of
Slater determinants to express $P\left|F\right>$ in the basis associated with $\left|F\right>$, and
as a result, $\left|\psi\right>$ can potentially describe strong correlations. There is however the
following pitfall. Since $\left|F\right>$ and $P\left|F\right>$ are associated with different static
scatterers, the orthogonality catastrophe generically causes the overlap
$\left<F\right|HP\left|F\right>$ to scale like $N^{-\alpha}$, where $N$ is the number of particles
in the system, and $\alpha\geq 0$ is determined by the phase shifts at the Fermi energy induced by
the static scatterers associated with $\left|F\right>$ and $P\left|F\right>$~\cite{Ohtaka}.
Generically then, $\left<F\right|HP\left|F\right>$ tends to zero in the thermodynamic limit. If
this happens, $\left|\psi\right>=\left|F\right>\pm P\left|F\right>$ is no better an approximation to
the true ground state than $\left|F\right>$ alone. However, the orthogonality catastrophe can 
be avoided by fine-tuning the
scattering phase shifts to produce $\alpha=0$. A natural way to do this is to associate a
non-interacting parent Hamiltonian $H_F$ with $\left|F\right>$, that has a similar form to the mean
field Hamiltonian. However, instead of determining the fields that are responsible for static
scattering by applying the mean field self-consistency condition, one views these fields as
variational parameters chosen to optimize $\left<\psi\right|
H\left|\psi\right>/\left<\psi\right|\left.\psi\right>$ for $\left|\psi\right>=\left|F\right>\pm
P\left|F\right>$. 

The same symmetry considerations were formulated early on within an approximate
variational treatment of the spin-boson model~\cite{Leggett}, known as the Silbey-Harris 
ansatz~\cite{Silbey,Harris}.
The spin-boson model describes dissipation in a quantum mechanical two-level system coupled to a bath of harmonic 
oscillators. After non-universal
ultraviolet modes are eliminated from the spin-boson model and the IRLM, the models can be mapped
onto each other~\cite{Guinea,Kotliar1996,Costi1999}. At weak dissipation, an ansatz of the form
$\left|\phi\right>=\left|+\right>\left|B_+\right>-\left|-\right>\left|B_-\right>$ accurately
approximates the ground state of the spin-boson model. Here $\left|\pm\right>$ refers to the state
of the two-level system, and $\left|B_\pm\right>$ describes a bath state in which each oscillator is
in the ground state corresponding to an equilibrium position that is shifted, with the shifts
depending on oscillator frequency and on the state of the two level system. It has subsequently been
shown that the Silbey-Harris ansatz can be systematically improved to arbitrary accuracy by forming
a linear combinations of {coherent} states of the form $\left|\phi\right>$ where the oscillator displacements
and weights are treated as variational parameters~\cite{Bera1,Bera2}. A natural 
generalization in the fermionic context results in an ansatz 
\begin{equation}
\left|\Psi\right>=\sum_{J=1}^M f_J (\left|F_J\right>\pm P \left|F_J\right>).\label{ansatz} 
\end{equation} 
The above variational state is a linear superposition of individual Slater pairs 
$\left|F_J\right>\pm P \left|F_J\right>$. In this way, a 
single ground state is associated with two or more non-interacting parent Hamiltonians $(H_{F,J},PH_{F,J}P)$. 
It is important to stress that the argument that we presented here is 
heuristic. Despite the equivalence between the spin-boson model and the IRLM, 
the ansatz (\ref{ansatz}) is not exactly equivalent to the ansatz employed previously for the
spin-boson model~\cite{Bera1,Bera2}.
Because a Slater pair describes all-fermionic microscopic degrees of freedom (including the
impurity), we can use Slater determinants $\left|F_J\right>$ in which the impurity and conduction band are hybridized.
In the spin-boson model, the microscopic degree of freedom associated with the impurity 
is a spin-$1/2$, while those of the bath are bosons, and one cannot construct Gaussian states that
hybridize them. Due to the presence of a hybridization channel in the IRLM, we may expect faster
convergence with respect to the number $M$ of Slater pairs, than was found for the sum
of Silbey-Harris terms in the spin-boson model. Indeed, below we find excellent agreement with
numerically exact IRLM results, in the challenging regime of strong correlations, for $M=2$. In contrast, it is not
uncommon to employ $M=8$ to reach good convergence in the same regime of the spin-boson
model~\cite{Bera1,Bera2}.

To assess the accuracy of our trial ground states, we compare to numerically exact results obtained
using NRG. Our model possesses an emergent energy scale, the Kondo temperature, that vanishes at
the quantum phase transition. As a result, a variational trial state can give a seemingly
reasonable approximation to the ground state energy, yet miss important features of the true ground
state at the Kondo scale. We therefore need to identify observable quantities that are sensitive
to the non-trivial correlations hosted by the IRLM. For this purpose, we demonstrate
the utility of the correlation matrix~\cite{Fishman,Debertolis} $Q$, and more particularly, its
eigenvalues. The correlation matrix can be defined as follows. Imagine viewing a
fermionic wave function as a state of a system of $N$ distinguishable particles that just happens to
be anti-symmetric under particle exchange. The correlation matrix is then $N$ times the reduced
density matrix of one of the $N$ particles. All its eigenvalues lie in the interval $[0,1]$. Owing
to the unavoidable entanglement between the hypothetical distinguishable particles that is implied
by antisymmetrization, $Q$ never describes a pure state. If (and only if) the fermionic wave
function is a pure Slater determinant, $Q$ has $N$ eigenvalues equal to one, and all others are zero.
The eigenvalues of $Q(1-Q)$ are therefore non-zero only when non-trivial many-body correlations are
present. To accurately reproduce the eigenvalues of $Q(1-Q)$ that are significantly different from
zero, requires that an approximate state accurately captures the many-body correlations of the true
ground state. In contrast to this, Hartree-Fock mean field theory approximates all eigenvalues of
$Q(1-Q)$ as zero.

The rest of this article is structured as follows. In Sec.~\ref{system} we formulate the model that
we study and present the discretized version used for numerics. We then provide the explicit form of
our ansatz and the associated parent Hamiltonians. At this point we also discuss the relation
between our approach and existing ones. In Sec.~\ref{results} we present numerical results. We
benchmark our ansatz against numerically exact results obtained using NRG, 
and compare to some existing variational approaches for fermionic impurity problems. Having
established the reliability of our method, we also present results in a regime inaccessible to NRG.
In Sec.~\ref{conclusion} we summarize our approach and main conclusions, and provide an outlook on
future work. Three appendices contain technical details about our calculations.

\section{System and ground state ansatz}
\label{system}

\subsection{Model and basic properties}
The IRLM describes spinless fermions in a crystal band with edges at $-t$ and $t$, interacting with
a localized orbital. We denote the annihilation operator of the localized orbital by 
$c_{-1}$. Band orbitals are labeled by their energy, and the associated annihilation operators are denoted $a_\varepsilon$.
We assume a constant density of states in the band.
The model incorporates tunnelling between the localized orbital and a single site (labeled $0$) of the crystal,
and a density-density interaction between the localized orbital and crystal site zero.
Tunnelling is controlled by the hybridization strength $\gamma$ and interactions by the coupling constant $U$. The
Hamiltonian reads:
\begin{eqnarray}
H_{\rm IRML}&=& \int_{-t}^t d\varepsilon\, \varepsilon a_\varepsilon^\dagger a_\varepsilon
+ U\left(n_{-1}-\frac{1}{2}\right)\left(n_0-\frac{1}{2}\right)\nonumber\\
&& +\frac{\gamma}{2\sqrt{t}}\int_{-t}^t d\varepsilon\, \left(c_{-1}^\dagger a_\varepsilon
+ a_\varepsilon^\dagger c_{-1}\right) \label{irlm}
\end{eqnarray}
where $n_{-1}=c_{-1}^\dagger c_{-1}$ and $n_0=c_0^\dagger c_0$ with
\begin{equation}
c_0=\frac{1}{\sqrt{2t}}\int_{-t}^t d\varepsilon\, a_\varepsilon.
\end{equation}
Throughout this article, we study the most interesting case, 
where the Fermi energy is aligned with the on-site energy $(=0)$ 
of the localized level, so that the system possesses particle-hole symmetry. The system then hosts a quantum phase transition. For $\gamma$ sufficiently smaller than $t$, the transition occurs
at $U_c\simeq -1.3 \,t$. 
For $U>U_c$, the system is in the symmetric phase, with a unique ground state satisfying the 
expectation value $\left<n_{-1}\right>=1/2$. For $U<U_c$ there are two degenerate ground states.
The phase transition is associated with spontaneous particle-hole symmetry breaking, and can be diagnosed 
by applying an infinitesimal on-site energy $b n_{-1}$ to the localized orbital.
In the symmetric phase, this produces only an infinitesimal change to $\left<n_{-1}\right>$. In the broken
symmetry phase on the other hand, $\left<n_{-1}\right>$ differs from 1/2 by a finite amount, even for infinitesimal $b$. 
Typical of quantum phase transitions, there is also
an emergent energy scale in the symmetric phase, that vanishes as the critical point is approached.
It is associated with the polarizability of the localized orbital and can be defined as 
\begin{equation}
T_{\rm K}=\frac{1}{4 \chi},
~\chi=\left.\partial_b\left<n_{-1}\right>\right|_{b=0}.\label{tk}
\end{equation}
As mentioned earlier, the IRLM can be mapped onto the Kondo model, and
$T_K$ is nothing but the Kondo temperature~\cite{Hanl}.

For numerical work, we have to truncate the above thermodynamic system to a finite set of
electronic modes. 
Because we want to make a direct comparison to NRG results, and because we
want to access long wavelengths at the lowest possible numerical cost, 
we will employ a logarithmic energy discretization:
\begin{equation}
\varepsilon_{n,\pm}=\pm\frac{1+\Lambda}{2\Lambda}\Lambda^{-n}t,~\mathrm{for}~n=0,1,2,\ldots,\Omega,
\end{equation}
with a discretization parameter $\Lambda>1$. The thermodynamic limit is recovered by sending $\Omega\to\infty$ followed
by $\Lambda\to 1$. While NRG becomes numerically too demanding for $\Lambda$ significantly less than $1.5$, it turns out
that many quantities reach values close to the thermodynamic limit for $\Lambda$ between $1.5$ and $2$. After
the logarithmic discretization of the energy, a standard tridiagonalization procedure maps the model onto a Wilson 
chain~\cite{Bulla2008} with Hamiltonian
\begin{eqnarray}
H&=&U\left(n_{-1} -\frac{1}{2}\right)\left(n_0 -\frac{1}{2}\right)
+\gamma\left(c_{-1}^\dagger c_0+c_0^\dagger c_{-1}\right)\nonumber\\
&&+\sum_{n=0}^{2\Omega-1} t_n\left(c_n^\dagger c_{n+1}+c_{n+1}^\dagger c_n\right).\label{chainh}
\end{eqnarray} 
Here, the operators $c_n$ with $n\geq 1$ are not associated with sites of the physical lattice, but
rather with energy shells corresponding to energy scales $\sim t \Lambda^{-n/2}$ above and below the
Fermi energy. Hopping between energy shells is controlled by an exponentially decaying hopping
amplitude
\begin{equation}
t_n=\frac{\left(1+\Lambda^{-1}\right)\left(1-\Lambda^{-n-1}\right)}{2\sqrt{1-\Lambda^{-2n-1}}\sqrt{1-\Lambda^{-2n-3}}}\Lambda^{-n/2} t.\label{deft}
\end{equation}
By truncating the chain to $2\Omega$ shells, one imposes an infrared cut-off of $\Lambda^{-\Omega} t$.
All the results we present are for the Hamiltonian (\ref{chainh}).

Apart from computing the ground state energy of the model, we will study the correlation matrix,
\begin{equation}
[Q]_{m,n}=\left<c_m^\dagger c_n\right>;\,m,\,n\in\{-1,0,\ldots,2\Omega\},\label{qmat}
\end{equation}
where the expectation value is with respect to the ground state.
The properties of $Q$ that were cited in the introduction can be derived by
considering the expectation value of an arbitrary additive single-particle operator 
\begin{equation}
\hat Z=\sum_{m,n=-1}^{2\Omega} Z_{mn}c_m^\dagger c_n.
\end{equation}
with respect to the $N$-particle ground state.
On the one hand $\left<\hat Z\right>={\rm Tr}\left(ZQ\right)$. On the other hand 
$\left<\hat Z\right>=N{\rm Tr}\left(Z\rho_1\right)$, where $\rho_1$ is the
reduced density matrix obtained by tracing out all but one particle. Since $Z$ is arbitrary, $Q=N\rho_1$.

The IRLM Wilson chain manifests particle-hole symmetry $H=P H P^\dagger$, where $P$ is the unitary and
hermitian particle-hole conjugation operator
\begin{equation}
P=\prod_{n=0}^\Omega \left(c_{2n}^\dagger+c_{2n}\right)\left(c_{2n-1}^\dagger-c_{2n-1}\right), 
\label{Pconj}
\end{equation}
with action:
\begin{eqnarray}
Pc_nP&=&(-1)^n c_n^\dagger,\nonumber\\
P\left|0\right>&=&c^\dagger_{2\Omega}\ldots c_{-1}^\dagger\left|0\right>
\end{eqnarray}

\subsection{Slater pair ansatz}
The approach outlined in the introduction then leads to a variational ground state ansatz of the form 
\begin{equation}
\left|\psi\right>=\sum_{J=1}^M f_J \left(\left|F_J\right>+\sigma P\left|F_J\right>\right).
\label{PairAnsatz}
\end{equation}
where $\sigma=\pm1$ is the eigenvalue of $\left|\psi\right>$ with respect to particle-hole conjugation $P$. 
Here $\left|F_J\right>$ is the Slater determinant ground state in the half-filled sector of the parent Hamiltonian
\begin{eqnarray}
H_J&=&\sum_{n=-1}^{2\Omega}\frac{\varepsilon_n^{(J)}}{\Lambda^{-n/2}} c_n^\dagger c_n
+\sum_{n=0}^{2\Omega}\frac{g_n^{(J)}}{\Lambda^{-n/2}} 
\left(c_{-1}^\dagger c_n+c_n^\dagger c_{-1}\right) \nonumber\\
&&+\sum_{n=0}^{2\Omega-1} t_n\left(c_n^\dagger c_{n+1}+c_{n+1}^\dagger c_n\right),
\label{Hparent}
\end{eqnarray}
For our definition of $P$, the ground state sector (of the symmetric phase) has $\sigma=-1$. 
The coefficients $f_J$, and the parameters $\varepsilon_n^{(J)}$ and $g_n^{(J)}$ are determined by minimizing 
\begin{equation}
E_{\rm var}=\frac{\left<\psi\right|H\left|\psi\right>}{\left<\psi\right|\left.\psi\right>}.
\end{equation}
In our definition of $H_J$, we multiply $\varepsilon_n^{(J)}$ and $g_n^{(J)}$ by scaling factors
$\Lambda^{n/2}$ appropriate for shell $n$. This produces a parameter space in which the region that
needs to be searched has roughly the same size in every direction. Technical details of the
variational calculation can be found in Appendices~\ref{apa} and \ref{apb}.

The parent Hamiltonians $H_J$ are generalizations of the Hartree-Fock mean field Hamiltonian
associated with the model. In the standard mean-field approach, only three effective
parameters appear, namely the two renormalized potentials $\epsilon_{-1} c_{-1}^\dagger
c_{-1}$ and $\epsilon_{0} c_{0}^\dagger c_{0}$ on the impurity and on site zero of the chain
respectively, and the renormalized hybridization $g_0 (c_{-1}^\dagger c_{0}+c_{0}^\dagger c_{-1})$
between the impurity and site zero. 
Indeed, minimizing the energy with respect to a single determinant, one
finds a minimum when $\varepsilon_n$ and $g_n$ are zero for $n\geq 1$, and the remaining
parameters obey the expected Hartree-Fock self-consistency conditions:
\begin{eqnarray}
\frac{\varepsilon_{-1}}{\sqrt{\Lambda}} &=& U\left(\left<n_0\right>-\frac{1}{2}\right),\\
\varepsilon_{0} &=& U\left(\left<n_{-1}\right>-\frac{1}{2}\right),\\
g_0 &=& \gamma-U\left<c_{-1}^\dagger c_0\right>.
\end{eqnarray}
This mean-field ansatz misses crucial Kondo physics.
Yet, building a Slater pair ansatz parametrized only with the three mean-field
parameters $\varepsilon_{-1}$, $\varepsilon_{0}$ and $g_0$, does not lower the energy. Indeed,
in the symmetric case where $\varepsilon_{-1}=\varepsilon_0=0$, the two members of the pair are
equivalent, and this ansatz reduces to the standard Hartree-Fock state. On the other hand, if the
parent Hartree-Fock state breaks particle-hole symmetry, it can be verified that the two members
of the pair are orthogonal to each other, due to the Anderson orthogonality catastrophe. Again, the variational energy of the Slater pair
does not improve with respect to standard mean-field theory. Thus, the 
long-range potential and hybridization in the parent Hamiltonian~(\ref{Hparent}) are crucial to capture Kondo correlations.
While more general Gaussian states \cite{AshidaA,AshidaB,Bravyi,Bauer} have been used recently, one of our
main goals is to show that restricting parametrization to a site dependent potential and
hybridization suffices to obtain accurate results for the IRLM.

\subsection{Other approaches}

To evaluate the quality of our trial state, it is useful to compare to existing methods based on 
comparable strategies. The most general of these is the one of
Ashida, Shi, Ba\~{n}uls, Cirac and Demler, which uses a canonical transformation to 
decouple the impurity from the bath~\cite{AshidaA,AshidaB}. The ground state of the transformed system 
is then approximated as a single Gaussian state~\cite{Kraus_2010,Weedbrook}. 
This approach does not aim to achieve arbitrary accuracy and 
only explores a restricted region of Hilbert space. Nonetheless, the 
error it makes for the ground state energy of the Kondo model is less than $0.5\%$ in a significant portion of the phase diagram. Another attractive feature is that it has proved extensible to the description
of dynamics.
We will refer to this method as the Canonically Transformed Gaussian approach (CTG).

To benchmark our approach, we applied 
the CTG approach directly to the IRLM. The appropriate canonical transformation 
for the Hamiltonian (\ref{chainh}) is:
\begin{eqnarray}
T &=& \frac{1}{\sqrt{2}}\left[1+P\left(2n_{-1}-1\right)\right],
\end{eqnarray}
with the particle-hole conjugation operator $P$ defined in Eq.~(\ref{Pconj}).
This transforms the conserved charge-conjugation parity into the occupation index of the localized
orbital, i.e. $TPT^\dagger=1-2n_{-1}$. In the symmetric phase and in the untransformed frame, the
unique ground state has charge conjugation parity $-1$. Hence, in the transformed frame, the
localized orbital is occupied. In the ground state sector where $n_{-1}=1$, the transformed Hamiltonian 
for sites $n=0,1,\ldots,2\Omega$ reads:
\begin{eqnarray}
THT^\dagger &=& \frac{U}{2}\left(n_0-\frac{1}{2}\right) +\gamma c_0 \mathcal P
 +\sum_{n=0}^{2\Omega-1} (t_n c_n^\dagger c_{n+1}+\mathrm{h.c.}) \nonumber\\
\label{THT}
\end{eqnarray} 
where 
\begin{equation}
\mathcal P=(c_{2\Omega}^\dagger+c_{2\Omega})(c_{2\Omega-1}^\dagger-c_{2\Omega-1})\ldots(c_0^\dagger+c_0)
\end{equation}
is the charge conjugation operator for the many-body system consisting of sites $0$ to $2\Omega$
such that $\mathcal P^\dagger=\mathcal P$, $\mathcal P^2=1$
and $\mathcal P c_n\mathcal P=(-1)^n c_n^\dagger$, $n=0,1,\ldots,2\Omega$. 
Note that the transformed Hamiltonian is completely non-local due to the term $\gamma
c_0 \mathcal{P}$.
In the transformed frame, the exact ground state takes the form 
$c_{-1}^\dagger\left|\Psi\right>$, where $\left|\Psi\right>$ is the many-body ground state of Eq.~(\ref{THT}).
The CTG approach assumes a trial state
$c_{-1}^\dagger\left|G\right>$, in which the Gaussian state $\left|G\right>$ is the ground state of a generic quadratic parent
Hamiltonian:
\begin{equation}
H_G=\sum_{mn=0}^{2\Omega} \left(h_{mn}c_m^\dagger c_n+\Delta_{mn}c_m^\dagger c_n^\dagger+\Delta_{mn}^*c_n c_m\right),
\end{equation}
with $n_{-1}\left|G\right>=0$.
Mapping back to the original frame, the CTG ansatz for the ground state of
$H$ in eq.\,(\ref{chainh}) then reads 
\begin{equation}
\left|\psi_{\rm CTG}\right>=\frac{1}{\sqrt{2}}\left(c_{-1}^\dagger \left|G\right>-\mathcal P\left|G\right>\right),\label{ctg}
\end{equation}
The CTG state is obtained by optimization over all possible parent Hamiltonians $H_G$. 

Two well-known earlier variational schemes can be obtained from the trial state (\ref{ctg}), by placing restrictions on the Gaussian state $\left|G\right>$.
When $\left|G\right>$ is taken to be of the form
\begin{equation}
\sum_{m=0}^{2\Omega} \psi_m c_m^\dagger \left|F\right>
\end{equation}
where $\left|F\right>$ is the $(\Omega-1)$-particle Fermi sea ground state of the Wilson chain (\ref{chainh}) with $U=0$, 
$\gamma=0$ and $n_{-1}=0$, one obtains the equivalent of Yosida's Kondo trial state~\cite{Yosida},
translated into the language of the IRLM.
Alternatively, when the parent Hamiltonian $H_G$ is taken to be of the form 
\begin{equation}
\sum_{n=0}^{2\Omega-1} t_n\left(c_n^\dagger c_{n+1}+c_{n+1}^\dagger c_n\right)+\sum_{n=0}^{2\Omega} v_n c_n^\dagger c_{n}
\end{equation}
i.e. the kinetic term of the Wilson chain, plus a particle-hole-symmetry breaking on-site energy $v_n$, one obtains
the IRLM equivalent of the original Silbey-Harris~\cite{Silbey,Harris} approximation for the Ohmic spin-boson model.
 
It is important to note that the CTG state does not contain terms of the form
$c_{-1}^\dagger c_{n}$ or $c_{n}^\dagger c_{-1}$, that hybridize the localized orbital with 
the rest of the chain, while our parent Hamiltonians do include such terms.
The omission of such terms in the CTG approach is a reasonable price to pay in order to have a formalism that applies to generic
systems in which the impurity may have very different degrees of freedom from the bath. However, in situations where the combined
impurity plus bath constitutes a system of indistinguishable particles, it seems reasonable to include hybridization terms.
For the IRLM, the hybridization terms allow the Ansatz to reduce to the exact ground state when $U=0$, and to be at least 
as accurate is Hartree-Fock mean field theory when $U\not=0$. On the other hand, the CTG parent Hamiltonian contains 
pairing terms $\Delta_{mn}c_m^\dagger c_n^\dagger+\Delta_{mn}^*c_n c_m$.
This may capture many-body correlations beyond the reach of a single Slater determinant. It is interesting to 
ask whether or not these terms mimic the effect of hybridization terms in our approach.
We also note that the number of variational parameters in the CTG approach scales quadratically 
with the system size, whereas in our approach it scales linearly.

\section{Numerical results}
\label{results}

\subsection{Ground state energy}

We now present numerical results in which we compare our Slater pair ansatz, optimized to yield the 
lowest possible variational energy $E_{\rm var}$, to various results: Hartree Fock mean field theory, the CTG 
approach, and NRG. The NRG results are very precise (9 or more significant digits), and are used to assess
the accuracy of approximate trial states. Unless otherwise stated, results are for the IRLM Wilson
chain with $\Lambda=1.5$, which allows good convergence to the thermodynamic limit of most
observable quantities. We chose the size parameter $\Omega=28$, which translates into a system of 
29 particles distributed among 58 orbitals and an infrared cut-off scale of $10^{-5}t$. We fixed the
hybridization to $\gamma=0.15 t$. The considerations that dictated this choice of parameters are
explained in Appendix \ref{pars}. With the parameters as chosen here, the Kondo length becomes
larger than the system size for $U<-0.9 t$. We are thus able to probe fully developed Kondo
correlations for interaction strengths $U>-0.9 t$. Our goal is not only to show that the trial state
yields a reasonable estimate for the ground state energy, but also that it accurately reproduces the
correlation structure encoded in the correlation matrix $Q$. To do so, we will compare NRG and 
variational results for the spectrum of $Q(1-Q)$.

\begin{figure}[h]
\begin{center}
\includegraphics[width=.95 \columnwidth]{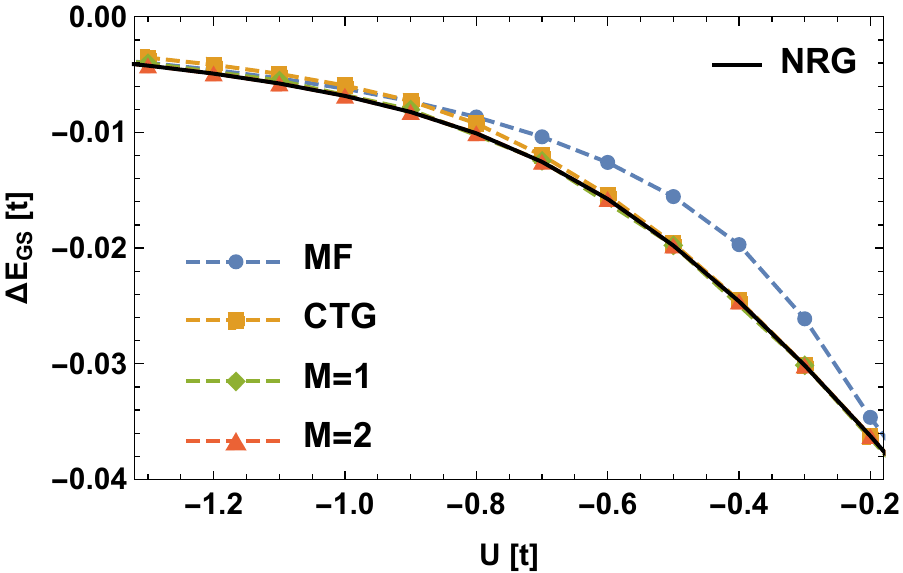}\\
\includegraphics[width=.95 \columnwidth]{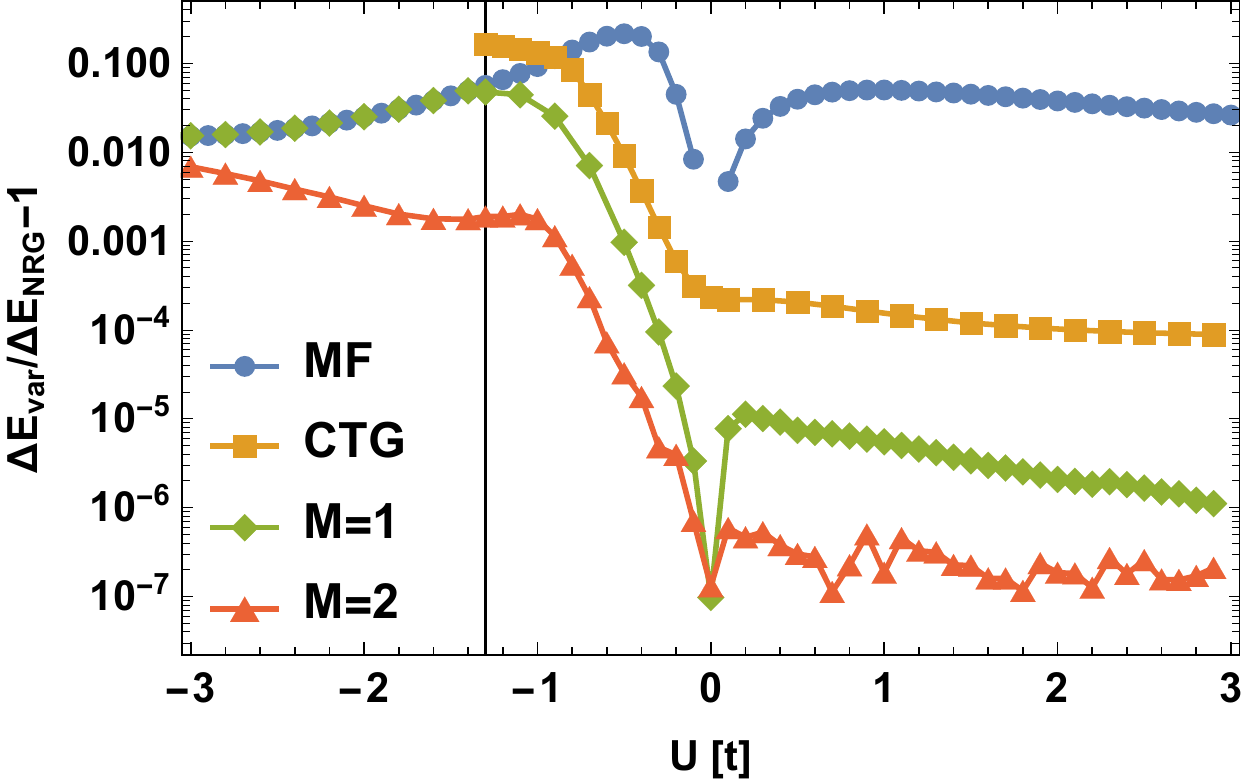}
\end{center}
\caption{Top panel: Relative ground state energy $\Delta E_{\rm GS}$ of the IRLM 
measured with respect to the $\gamma=0$ ground state energy, as a function of interaction 
strength $U$. Results are for the Wilson chain with $\Omega=28$ sites, discretization parameter
$\Lambda=1.5$ and hybridization $\gamma=0.15 t$.
We compare the essentially exact NRG results, to Hartree-Fock mean field theory (MF),
the CTG approach, and the $M=1$ and $M=2$ Slater pair approximations. 
Bottom panel: Relative error $\Delta E_{\rm var}/\Delta E_{\rm NRG}-1$ with respect to the NRG
computation, shown in log scale and for a larger region of the $U$-axis than in the top panel.
\label{f2}}
\end{figure}

The first variational results we present are for the optimized energy $E_{\rm var}$ as a function of $U$.
In Fig.~\ref{f2} we compare the different variational approaches with NRG. Here are shown results for
Hartree-Fock mean field theory, the CTG approach, and the $M=1$ and $M=2$ Slater pair
approximations. For a given value of $U$, we compute the relative ground state energy 
$\Delta E_\mathrm{GS}$, with respect to the ground state energy of the corresponding system with $\gamma=0$.
This subtracts a large kinetic energy contribution of the particles in the chain as well as a trivial 
contribution $\sim -|U|/4$ at large $|U|$. We denote energies measured from this off-set as 
$\Delta E_{\rm var}$ for the variational states and $\Delta E_{\rm NRG}$ for the NRG benchmark. 
Both decrease monotonically as a function of $U$, from $0$ at large negative $U$, to $-|\gamma|$
at large positive $U$.

Comparing the Slater pair approximation to the numerically exact NRG calculation and Hartree-Fock,
we see the following. In the phase with broken particle-hole symmetry ($U<-1.3 t$), the $M=1$ ansatz is equivalent to
Hartree-Fock mean field theory, whereas in the symmetric phase it clearly outperforms the latter.
Indeed Hartree-Fock predicts spontaneous symmetry breaking 
(non-zero value 
of $\big<n_{-1}\big>-1/2$)
already for $U<-0.3t$.
In the interval $-1.3 t<U<0$, the relative error associated with
the $M=1$ ansatz drops rapidly from $0.05$ of the ground state energy, to zero. For $U=-0.9t$,
where Kondo correlations become fully developed, the $M=1$ ansatz produces a relative error of less
than one percent. The $M=1$ ansatz makes an error that is typically five times smaller than the CTG
approach in the interval $-1.3 t<U<0$, while involving far fewer variational parameters.
For positive $U$, the relative error is at most $10^{-5}$ of the ground state energy, dropping to 
$10^{-6}$ at $U=3t$. The error is between $20$ and $100$ times smaller than that associated with 
the CTG approach, showing that the Slater pair ansatz embodies a better representation of
the physics of the IRLM ground state.

Improving our ansatz by using $M=2$ Slater pairs further lowers the ground state energy. In the
symmetry broken phase for $U=-2t$, the error is ten times smaller than Hartree Fock, and the
absolute error is $4\times 10^{-6}t$. At the phase transition, the relative error is $5\times
10^{-3}$ of the ground state energy, or $20$ times smaller than Hartree Fock. The relative error at
$U=-0.9t$, where Kondo correlations are fully developed, is one part in $1000$, two orders of
magnitude more accurate than the CTG approach. For larger $U$, the error rapidly drops further. For
$-0.9 t<U<0$, the $M=2$ ansatz is typically between 2 and 4 orders of magnitude more accurate than
Hartree-Fock, and 2 orders of magnitude more accurate than the CTG approach. For $U>0$ the maximum
relative error is $10^{-6}$ of the ground state energy, and drops to $10^{-7}$ at $U=3t$. This
is about 3 orders of magnitude more accurate than the CTG approach.

Examining the CTG approach further, we find that there is no pairing in the region $U>-0.9 t$ where
Kondo correlations are fully developed. In other words, in this regime, $\left|G\right>$ reduces to
an $\Omega$-particle Fermi sea (Slater determinant). Thus we conclude that pairing correlations
allowed in the CTG approach are not able to mimic the hybridization terms included in the parent
Hamiltonians of the Slater pair approximation. We stress that this conclusion is for the IRLM.
Although the IRLM is equivalent to the Kondo model, applying the CTG approach to the Kondo model is
in principle not equivalent to applying it to the IRLM. 
Since the fermions in the IRLM are ``squares" of the fermions in the Kondo model, the degrees of 
freedom that are assigned a Gaussian correlation structure are not the same in the two cases. 
Nonetheless, in their application of the CTG approach to the Kondo model, Ashida and co-workers find 
similar relative errors as we do for the IRLM, ranging from around $0.1$ close to the phase transition, 
to $\sim 10^{-4}$ deep in the symmetric phase. We do not show CTG results for the symmetry broken phase 
of the IRLM. In this phase, the restrictions placed on the CTG trial state conspire with the orthogonality 
catastrophe to produce a result that is guaranteed to be worse than Hartree Fock mean field theory. It 
should be noted that in contrast, when the CTG approach is applied to the Kondo model, the ferromagnetic
phase, equivalent to the symmetry-broken phase of the IRLM, yields the most accurate results,
indicating that CTG performs better for the Kondo model than for the IRLM, presumably due to
the fact that the Kondo impurity spin is distinguishable from the bath electronic states.

\subsection{Correlation matrix spectrum}
We have demonstrated that, unlike mean field theory, or the CTG approach, the $M=2$ Slater pair
ansatz is reliable throughout the whole phase diagram of the IRLM, and is nearly exact for $U>0$. We now 
focus on the strongly correlated physics of the symmetric phase, $-0.9t<U<0$ for the chosen parameters
of the model. Our aim is to further quantify the extent to
which our variational approach reproduces the many-body correlations present in the ground state.
For this purpose, we consider the correlation matrix $Q$ (\ref{qmat}). Before presenting results for
our trial states, we review a few relevant properties of $Q$ spectrum. In the symmetric phase,
particle-hole symmetry implies that the eigenvalues of $Q(1-Q)$ are at least two-fold degenerate. 
When representing the correlation spectrum, we will show only one member of each pair and order 
them in decreasing order $\lambda_1\geq \lambda_2 \geq \ldots \geq \lambda_{\Omega+1}$. NRG results 
reveal a further approximate two-fold
degeneracy of the largest eigenvalue of $Q(1-Q)$, i.e. $\lambda_1\simeq \lambda_2$. Exponential
decay abruptly sets in from $\lambda_3$, i.e. $\lambda_n=A e^{-x n}$ for $n\geq 3$. The approximate
four-fold degeneracy of the largest eigenvalues of $Q(1-Q)$ reveals a Bell-state like nature of the
IRLM ground state. This is related to the fact that at negative $U$, the localized orbital and the site 
zero of the chain tend to be either both filled or both empty (similarly at positive $U$, if the localized orbital
is filled, the site zero tends to be empty, and vice versa).

\begin{figure}[h]
\begin{center}
\includegraphics[width=.99 \columnwidth]{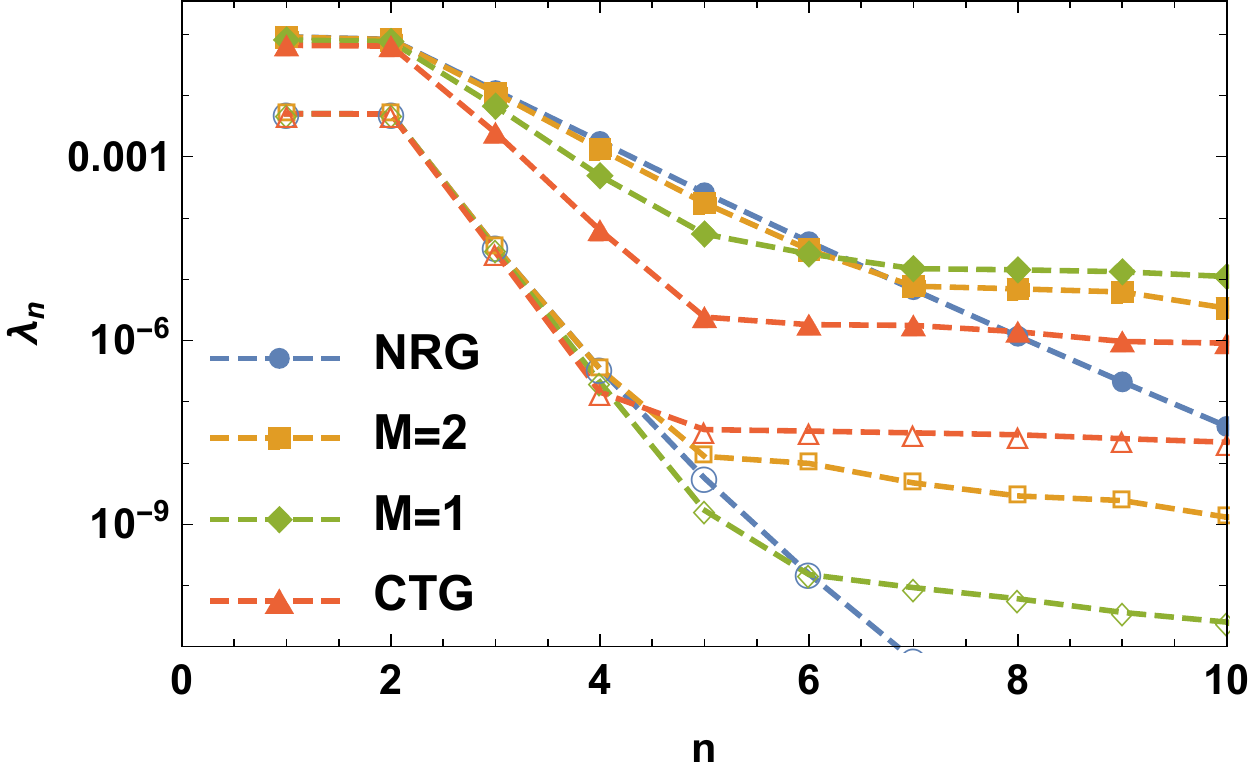}\\
\includegraphics[width=.99 \columnwidth]{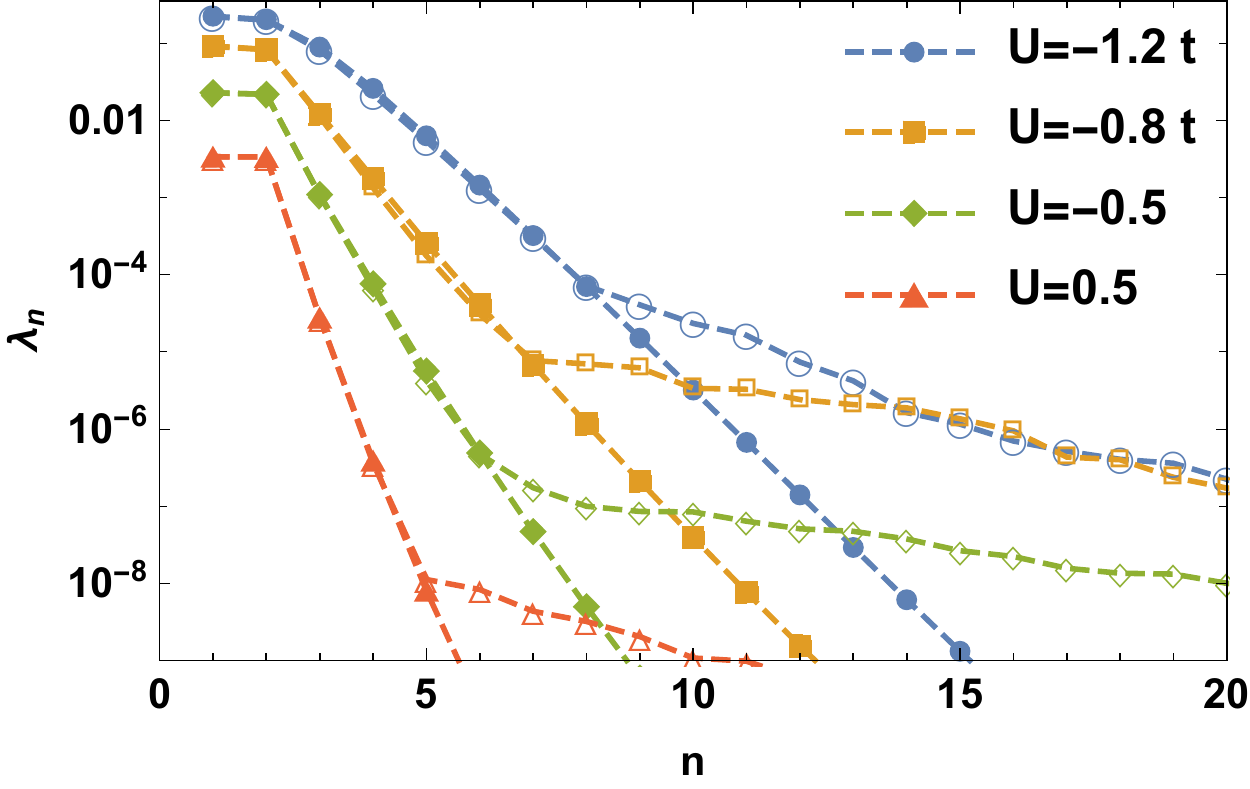}
\end{center}
\caption{Eigenvalue spectrum of $Q(1-Q)$. Top panel: Comparison between NRG, the CTG approach and
the $M=1$ and $M=2$ Slater pair approximations for $U=-0.6t$ (filled symbols) and $U=0.8t$ (open
symbols). Results are for $\gamma=0.15t$, $\Lambda=1.5$ and $\Omega=28$. Bottom panel: Comparison 
between NRG and the $M=2$ Slater pair ansatz, at various $U$, for the same set of parameters.
\label{f3}}
\end{figure}

In Fig.\,\ref{f3} we compare our variational results to NRG for the spectrum of $Q(1-Q)$. In the top
panel, we present results for the CTG, and $M=1$ and $M=2$ Slater pair approximations. We see that
all three variational states produce good results for $\lambda_1$ and $\lambda_2$, the two largest
eigenvalues that are most directly linked to the Bell-like nature of the particle-hole symmetric
ground state. At $U=-0.8t$ (filled symbols), it is clear that eigenvalues $\lambda_3$ to $\lambda_5$
are underestimated more severely, the less accurate the variational state, i.e. the CTG approach
gives the smallest eigenvalues, followed by $M=1$, and then $M=2$, which very nearly coincides with
NRG. Recalling that all eigenvalues of $Q(1-Q)$ are zero for a single Slater determinant, the
interpretation is as follows: The CTG approach and the $M=1$ ansatz underestimate eigenvalues
$\lambda_3$ to $\lambda_5$ because these trial states are less correlated than the true ground
state. We further see that beyond a certain index $n$, all variational states produce spurious
plateaus in the spectra, whereas the true spectrum continues to decay exponentially with increasing
$n$. These plateaus arise because a given family of variational states can only produce physical
correlations above a certain resolution. Weaker correlations present in the variational state are
determined not by physical effects, but by the limited form imposed on the ansatz. For
$U=-0.8t$ we see that the $M=2$ ansatz accurately reproduces $\lambda_1$ to $\lambda_7$, thus
accounting for all eigenvalues down to $10^{-5}$. At $U=0.8t$ (open symbols), all three trial states
reproduce the spectrum of $Q(1-Q)$ well, down to eigenvalues $\sim 10^{-7}$. This is consistent with
the increased accuracy of the ground state energy at positive $U$ with respect to negative $U$,
as seen in Fig. \ref{f2}. Not only do the eigenvalues decay faster for positive $U$, but their
magnitude is also smaller for positive $U$ than for negative $U$. Thus, the improved accuracy of all
the variational states for positive $U$ is tied to the fact that many-body correlations are weaker
in this parameter regime. In the lower panel of Fig.\,\ref{f3}, we compare the
$M=2$ ansatz (open symbols) to NRG at various values of $U$. At $U=-1.2 t$, the first eight
eigenvalues are accurately reproduced, representing a threshold between $10^{-4}$ and $10^{-5}$.
This threshold improves to $10^{-8}$ at $U=0.5t$. However, at $U=0.5t$ the ground state contains
weaker many-body correlations than close to the phase transition and as a result, there are only
five eigenvalues above the threshold to the spurious plateau behavior.

\begin{figure}[h]
\begin{center}
\includegraphics[width=.99 \columnwidth]{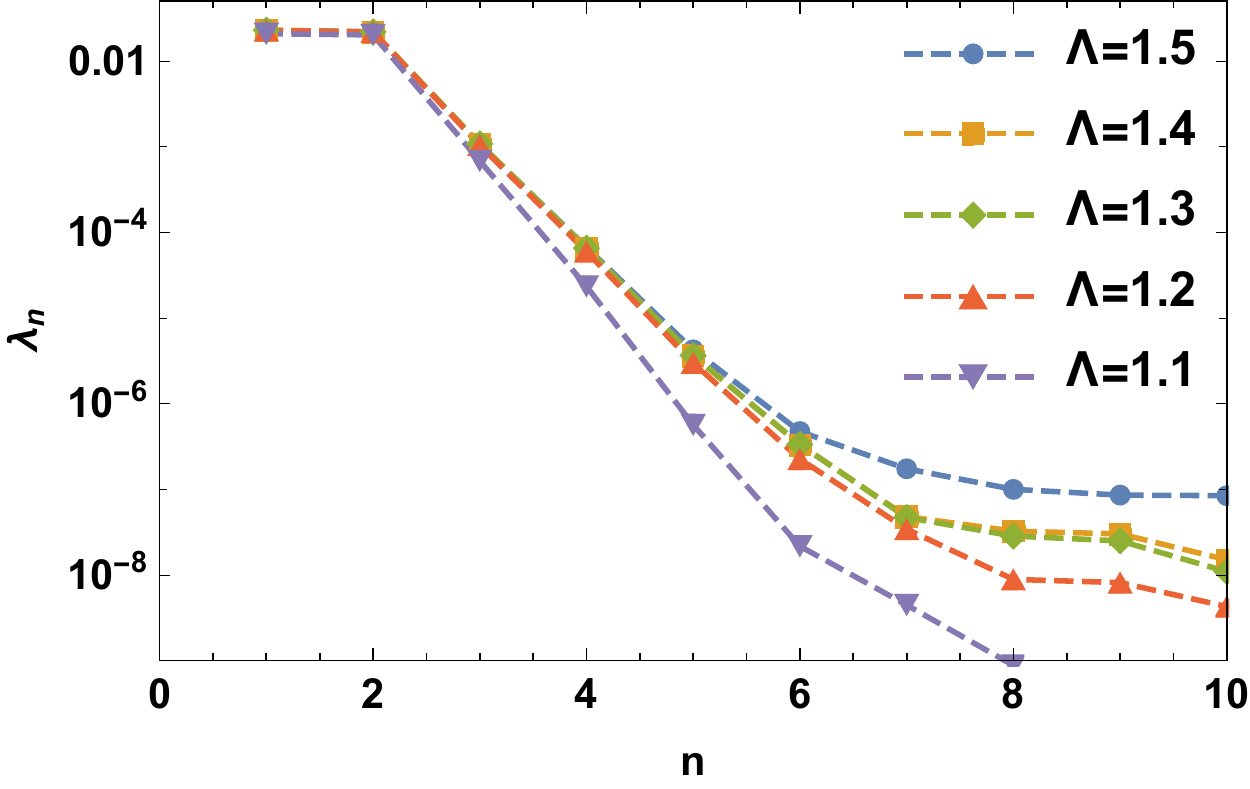}
\end{center}
\caption{Eigenvalue spectrum of $Q(1-Q)$ in descending order, for various discretization parameters 
$\Lambda$, for $U=-0.5\,t$, $\gamma=0.15\,t$, and $\Omega=28$. Only one eigenvalue of each degenerate pair 
is plotted. Results were obtained variationally using $M=2$ pairs of Slater determinants. The infrared scale
$\Lambda^{-\Omega}\,t$ varies from $5\times 10^{-6}\,t$ at $\Lambda=1.5$ to $7\times10^{-2} t$
at $\Lambda=1.1$, while the Kondo temperature is $3\times 10^{-3}\,t$. 
The apparent weakening of
correlations for $\Lambda=1.1$ is thus due to finite size effects that cut off the full development
of the Kondo state.
\label{f5}}
\end{figure}

It is interesting to note that the Slater pair ansatz remains accurate beyond the range of $\Lambda$
accessible through NRG. Using NRG
we have verified that the spectrum of $Q(1-Q)$ has no discernible $\Lambda$ dependence for
$2>\Lambda>1.5$~\cite{Debertolis}. This provides strong evidence that $\Lambda=1.5$ results have converged to the
continuum limit (\ref{irlm}). However, it is challenging to push the NRG calculation of $Q$ well beyond
$\Lambda=1.5$. In contrast to this, we could perform $M=2$ variational calculations at $U=-0.5\,t$,
$\gamma=0.15\,t$, $\Omega=28$, and successively lower $\Lambda$. Results are presented in
Fig.\,\ref{f5}. We see that the part of the $Q(1-Q)$ spectrum that is accurately reproduced by the
ansatz is insensitive to $\Lambda$, for $\Lambda\geq 1.2$, but changes significantly for
$\Lambda=1.1$. To understand this, we note that the Kondo temperature $T_K\sim 3\times 10^{-3} t$ 
is of the same order as the infrared cut-off for $\Lambda=1.2$, and an order of magnitude smaller 
than the infrared cut-off when $\Lambda=1.1$. (The number of sites on the Wilson chain was kept fixed when
changing the $\Lambda$ parameter). We conclude that the spectrum of $Q(1-Q)$ is insensitive to
$\Lambda$ as long as the Kondo temperature is larger than the infrared cut-off. We therefore expect
the spectrum of $Q(1-Q)$ to remain invariant if we take the continuum limit by first sending the
infrared cut-off to zero ($\Omega\to\infty$) before sending $\Lambda\to1$. This confirms
that the 
exponential decay of eigenvalues is not an artefact of the logarithmic discretisation of the conduction 
band, and is an intrinsic property of the quantum impurity problem~\cite{Bravyi,Debertolis}.

\section{Discussion and conclusions}
\label{conclusion}

Quantum impurity problems in which a localized orbital hosts electrons that interact with a
conduction band, play an important role in the field of strongly correlated electrons. Here we
investigated the wavefunction of the simplest, spinless case known as the interacting resonant
level model, a system that also provides a window into the Kondo problem. We formulated a
variational ansatz that provides an intuitive picture of the structure of the correlated ground
state. Our rational was to restore a
discrete symmetry that is spuriously broken in mean field theory, by forming appropriate linear
combinations of symmetry broken mean field states. The key insight is that this must be done in a
way that avoids the orthogonality catastrophe between symmetry broken terms. 
Thus we arrived at a
state that is a linear combination of non-orthogonal Slater determinants,
 each associated with a
different static scatterer in place of the dynamic impurity. Much like mean field theory, the
natural variational parameters are the matrix elements of the associated single particle
Hamiltonians. Focussing on the Hartree and Fock channels, we endowed each parent
Hamiltonian with a set of variational parameters whose size scales linearly with the size of the
system. We were able to obtain quantitatively accurate results up to very large correlation (Kondo) lengths. 
We compared our method to existing variational approaches that construct
correlated states using uncorrelated electronic states as building blocks. 
The comparison to the canonically transformed Gaussian
approximation~\cite{AshidaA,AshidaB} showed that pairing-type correlations, allowed by generic Gaussian states, are not
relevant for the physics of the IRLM. Clarifying for which families of impurity models such off-diagonal
correlations play a major role would be physically insightful.

We find that our method is significantly more accurate than comparable methods that do not
harness the full power of the superposition principle. Indeed, it has recently been shown that superpositions of Gaussian states can approximate
the ground states of fermionic impurity systems to any desired accuracy using resources that
scale quasi-polynomially in $1/\mbox{accuracy}$ \cite{Bravyi}. Our result demonstrates that physical insight into the specific problem under consideration can lead to practical algorithms employing the superposition principle, for which the number
of variational parameters are $\mathcal O(\mbox{system-size})$. This may significantly
simplify the optimization problem compared
to existing algorithms that use a general parametrization of Gaussian states, involving
$\mathcal O(\mbox{system-size}^2)$ variational parameters.

By studying the eigenvalues of the correlation matrix $Q$, we were also able
to quantify the extent to which our trial state encodes many-body correlations correctly. Requiring
the ansatz to reproduce the spectrum of $Q(1-Q)$ is a far more stringent and unbiased criterium than
reproducing expectation values for a small set of observables, such as energy and the order
parameter. Again, the conclusion is that our ansatz does an excellent job even at very large correlation
lengths. As the critical interaction strength $U_c$ is approached from within the symmetric phase, the 
ground state becomes more and more correlated. (The spectrum of $Q$ has more and more eigenvalues
significantly different from zero or one.) It is interesting to note the significant improvement in
accuracy that is obtained in this regime when a second Slater pair ($M=2$) is added to the single
pair ($M=1$) ansatz. It implies that the introduction of new Slater pairs mirrors the build-up of
correlations as the critical point is approached. We envision future work in which we develop
numerical minimization methods that can efficiently explore the larger parameter space associated
with more Slater pairs, and in this way quantify the above statement more precisely. It would for
instance be insightful to know how many Slater pairs are required for a specified accuracy, as a
function of the distance from the critical point, and to relate it to the bound derived in \cite{Bravyi}.

\acknowledgments We thank the National Research Foundation of South Africa (Grant No. 90657), and 
the CNRS PICS contract FERMICATS for support.

\appendix

\section{Matrix elements involving non-orthogonal Slater determinants}
\label{apa}
To evaluate the matrix elements of the correlation matrix $Q$, and the expectation value of the
Hamiltonian with respect to our trial state, we have to compute quantities of the form
$\left<X\right| \hat O\left|Y\right>$ where $\hat O$ is a product of up to four creation and
annihilation operators from the set $\{c_m;\,c_m^\dagger|\,m=-1,0,\ldots,2\Omega\}$, while
$\left|X\right>$ and $\left|Y\right>$ are single Slater determinants. What makes the situation
slightly unusual is that the single particle orbitals that are naturally associated with
$\left|X\right>$ and $\left|Y\right>$ are drawn from distinct single particle bases, and hence the
orbitals of $\left|X\right>$ are not the same or orthogonal to those of $\left|Y\right>$. In this
appendix we derive simple formulas applicable to this situation. We start by defining the objects we
need.

Let $N\leq 2\Omega+2$ be the number of particles in the system. Let $\{x_\alpha|\alpha=1,\ldots,N\}$
and $\{y_\alpha|\alpha=1,\ldots,N\}$ be two sets of fermion annihilation operators. The members of
each set can be expressed as linear combinations of $c_{-1},\,\ldots,\,c_{2\Omega}$. We denote the
respective expansion coefficients $X_{j\alpha}$ and $Y_{j\alpha}$,
which are rectangular $(2\Omega+2)\times N$ matrices. Throughout this section we will
use Einstein summation convention to imply sums over repeated indices. We can thus write 
\begin{equation}
x_\alpha^\dagger=c_j^\dagger X_{j\alpha},~y_\alpha^\dagger=c_j^\dagger Y_{j\alpha}.
\end{equation}
Throughout, greek indices imply a range $\{1;\ldots;N\}$ over particles, while lower case roman indices 
imply a range $\{-1;\ldots;2\Omega\}$ over orbitals. Note that in contrast to the original fermions $c_m$, 
neither the $x_\alpha$ 
nor the $y_\alpha$ operators are associated with complete single particle bases. In what follows below, we 
do not even need to assume that the orbitals associated with $\{x_\alpha|\alpha=1,\ldots,N\}$ are mutually 
orthogonal, only that they are linearly independent. (The same goes for $\{y_\alpha|\alpha=1,\ldots,N\}$.) 
We assume that the $x_\alpha$ and $y_\alpha$ operators are members of different single particle basis,
so that
\begin{equation}
\{x_\alpha,y_\beta^\dagger\}=M_{\alpha\beta},~M=X^\dagger Y.
\end{equation}
Using our two sets of $N$ creation operators, we construct two single-Slater determinants
\begin{equation}
\left|X\right>=x_N^\dagger\ldots x_1^\dagger\left|0\right>,~\left|Y\right>
=y_N^\dagger\ldots y_1^\dagger\left|0\right>.
\end{equation}
The overlap between $\left|X\right>$ and $\left|Y\right>$ is therefore
\begin{equation}
\left<X\right|\left.Y\right>={\rm Det}\,M.
\end{equation}
In the formulas we present below, we assume ${\rm Det}\,M\not=0$, so that $M^{-1}$ exists. However,
the limit ${\rm Det}\,M\to 0$ is regular.

First, we consider an arbitrary single-particle additive operator
\begin{equation}
\hat Z=Z_{mn}c_m^\dagger c_n\label{spao}.
\end{equation}
We will prove that its overlap between two distinct Slater determinants
can be evaluated as:
\begin{equation}
\left<X\right|\hat Z\left| Y\right>
=\left<X\right|\left.Y\right>{\rm Tr}\left[X^\dagger Z Y M^{-1}\right].\label{2ferm} 
\end{equation}

For the purpose of the proof, we define new fermion creation operators, and an associated Slater determinant 
\begin{equation}
\bar y_\alpha^\dagger(\lambda)=e^{\lambda \hat Z}y_\alpha^\dagger e^{-\lambda \hat Z}
=c_j^\dagger \left[e^{\lambda Z}Y\right]_{j\alpha},
~ \left|\bar Y(\lambda)\right>=e^{\lambda \hat Z}\left|Y\right>.
\end{equation}
The anti-commutator between $x_\alpha$ and $\bar y_\beta^\dagger(\lambda)$ evaluates to
\begin{equation}
\left\{x_\alpha,\bar y_\beta^\dagger(\lambda)\right\}
=\bar M(\lambda)_{\alpha\beta},~\bar M(\lambda)=X^\dagger e^{\lambda Z}Y, 
\end{equation}
and the overlap between $\left|X\right>$ and $\left|\bar Y(\lambda)\right>$ gives
\begin{equation}
\left<X\right|\left.\bar Y(\lambda)\right>={\rm Det}\,\bar M(\lambda).
\end{equation}

The result we want to prove now follows by noting that $\left<X\right|\left.\bar Y(\lambda)\right>$
can be used as a generating function for
$\left<X\right|\hat Z\left| Y\right>$ i.e.
\begin{eqnarray}
\left<X\right|\hat Z\left|Y\right>&=&\partial_\lambda \left. \left<X\right|\left.\bar Y(\lambda)\right>\right|_{\lambda=0}\nonumber\\
&=&\left.\partial_\lambda \exp{\rm Tr}\,\ln \bar M(\lambda)\right|_{\lambda=0}\nonumber\\ 
&=&\left.{\rm Det}\,\bar M(\lambda){\rm Tr}\,\left\{\left[\partial_\lambda \bar M(\lambda)\right]
\bar M(\lambda)^{-1}\right\}\right|_{\lambda=0}\nonumber\\
&=& \left<X\right|\left.Y\right>{\rm Tr}\left[X^\dagger Z Y M^{-1}\right],
\end{eqnarray}
 which completes the proof.
 
To compute the expectation value of interaction terms, we also need to evaluate 
quantities of the form $\left<X\right|\hat Z_1 \hat Z_2\left|Y\right>$
where both $\hat Z_1$ and $\hat Z_2$ are single-particle additive operators of the form
(\ref{spao}). We do so employing a strategy that is similar to the above. We use a generation function
\begin{equation}
\left<X\right|e^{\lambda_1 \hat Z_1} e^{\lambda_2 \hat Z_2}\left|Y\right>
={\rm Det}\left(X^\dagger e^{\lambda_1 Z_1}e^{\lambda_2 Z_2}Y\right), 
 \end{equation}
such that 
\begin{equation}
\left<X\right|\hat Z_1 \hat Z_2\left|Y\right>
=\left.\partial_{\lambda_2}\partial_{\lambda_1}\left<X\right|e^{\lambda_1 \hat Z_1} e^{\lambda_2 \hat Z_2}\left|Y\right>
\right|_{\lambda_1=\lambda_2=0}.
\end{equation}
On the right hand side we first take the $\lambda_1$ derivative and subsequently set $\lambda_1$ to zero, to obtain 
\begin{eqnarray}
&&\left.\partial_{\lambda_1}\left<X\right|e^{\lambda_1 \hat Z_1} e^{\lambda_2 \hat Z_2}\left|Y\right>
\right|_{\lambda_1=0}\nonumber\\
&=&\left<X\right|e^{\lambda_2 \hat Z_2}\left|Y\right>{\rm Tr}
\left[X^\dagger Z_1 e^{\lambda_2 Z_2} Y\left(X^\dagger e^{\lambda_2 Z_2} Y\right)^{-1}\right].\nonumber\\
\end{eqnarray}
Then taking the $\lambda_2$ derivative and setting $\lambda_2$ to zero, we arrive at the final result
\begin{align}
&\left<X\right|\hat Z_1 \hat Z_2\left|Y\right>
=\left<X\right| \hat Z_1\left|Y\right>\left<X\right| \hat Z_2\left|Y\right>/\left<X\right|\left.Y\right>\nonumber\\
&+\left<X\right|\left.Y\right>{\rm Tr}\left[X^\dagger Z_1(1-YM^{-1}X^\dagger) Z_2 Y M^{-1}\right].\label{4ferm} 
\end{align}

While the above form will be most useful for our numerical calculations, further insight into the
result can be gained by defining four new fermion annihilation operators $q_1,\,\ldots,q_4$, and 
two single particle additive operators $q_1^\dagger q_2$ and $q_3^\dagger q_4$, i.e. 
\begin{equation}
q_{i}=c_j A_{ji},~ \hat Z_1=q_{1}^\dagger q_{2},~\hat Z_2=q_{3}^\dagger q_{4}. 
\end{equation}
Substituting this $\hat Z_1$ and $\hat Z_2$ into the result (\ref{4ferm}) and commuting $q_2$ past 
$q_3^\dagger q_4$, we obtain
\begin{align}
&\left<X\right| q_1^\dagger q_2^\dagger q_4 q_3\left|Y\right>
=\frac{1}{\left<X\right|\left.Y\right>}\Big[\left<X\right|q_1^\dagger q_3\left|Y\right>
\left<X\right|q_2^\dagger q_4\left|Y\right>\nonumber\\
&-\left<X\right|q_1^\dagger q_4\left|Y\right>
\left<X\right|q_2^\dagger q_3\left|Y\right>\Big],\label{wick}
\end{align}
which is a very straight-forward generalization~\cite{Lowdin} of the familiar 
Wick's theorem when $\left|X\right>=\left|Y\right>$.
 
\section{Variation of the energy}
\label{apb}
We used a quasi-Newton method to optimize trial states. We thus had to calculate the expectation value 
\begin{equation}
\left<E\right>=\frac{\left<\psi\right|H\left|\psi\right>}{\left<\psi\right|\left.\psi\right>},
\end{equation}
of the Hamiltonian with respect to the trial state, as well as its variation $\delta\left<E\right>$
in response to changes of the variational parameters $\{f_J,\varepsilon^{(J)}_n,g^{(J)}_n\}$
appearing in the set of parent Hamiltonians~(\ref{Hparent}).
The expectation value $\left<E\right>$ as well as its $f_F$ derivatives can be calculated directly from
the results of the previous section. The derivatives with respect to $\varepsilon^{(J)}_n$ and
$g^{(J)}_n$ requires some further analysis, which we present here.
 
Let $\delta$ stand for the partial derivative with respect to any of the $\varepsilon^{(J)}_n$ or
$g^{(J)}_n$. Note that the IRLM Hamiltonian as well as the parent Hamiltonians associated with our
trial state can be simultaneously represented as real symmetric matrices. We can therefore perform
our analysis in a real rather than complex Hilbert space, and this allows us to write the variation
of $\left<E\right>$ as 
\begin{equation}
\delta\left<E\right> =2\frac{\left<\delta\psi\right|H\left|\psi\right>-\left<E\right>
\left<\delta\psi\right|\left.\psi\right>}{\left<\psi\right|\left.\psi\right>}.
\end{equation}

Recall that our trial state is
\begin{equation}
\left|\psi\right>=\sum_{J=1}^M f_J (1-P)\left|F_J\right>,
\end{equation}
where $\left|F_J\right>$ is the Fermi-sea ground state of the parent Hamiltonian $H_J$, and $P$
the conjugation operator over all fermions. The partial
derivative of the trial state is therefore related to the partial derivative $\delta H_J$ of the
parent Hamiltonian by first order perturbation theory on $H_J$. We note that $\delta H_J$ is a
single-particle additive operator that creates one particle-hole pair in the Fermi sea
$\left|F_J\right>$. We denote by $x_m$ ($m=1,\ldots,2\Omega+2$) the fermion annihilation 
operators associated with the complete single-particle basis in which the given $H_J$ that we are 
varying is diagonal. When referring to operators associated with the $\Omega+1$ lowest energy
orbitals of $H_J$ that are occupied, we use an unprimed greek index, i.e. $x_\alpha$,
$\alpha=1,\ldots,\Omega+1$. When referring to one of the $\Omega+1$ highest energy orbitals of
$H_J$, that are unoccupied, we use a primed greek index, i.e. 
$x_{\alpha'}$, $\alpha'=\Omega+2,\ldots,2\Omega$. With these conventions we then have
\begin{eqnarray} 
\left|\delta\psi\right>&=&f_J\sum_{\alpha=1}^{\Omega+1}\sum_{\beta'=\Omega+2}^{2\Omega} 
x_{\beta'}^\dagger x_\alpha \left|F_J\right>\Delta_{\alpha{\beta'}}\nonumber\\
\Delta_{\alpha\beta'}&=&\frac{\left<F_J\right|x_\alpha^\dagger x_{\beta'} 
\delta H_J\left|F_J\right>}{\varepsilon_\alpha^{(J)}-\varepsilon_{\beta'}^{(J)}},
\end{eqnarray}
and hence
\begin{align}
\delta\left<E\right>=&4f_J\sum_{J'=1}^M\sum_{\alpha=1}^{\Omega+1}\sum_{{\beta'}=\Omega+2}^{2\Omega}\Delta_{\alpha\beta'} 
a_{J'} O_{J'\alpha\beta'}\nonumber\\
 O_{J'\alpha\beta'}=&\frac{\left<F_J\right|x_\alpha^\dagger x_{\beta'} 
H\left|F_{J'}\right>-\left<F_J\right|x_\alpha^\dagger x_{\beta'} (H-\left<E\right>)\left| \tilde F_{J'}\right>}
{\left<\psi\right|\left.\psi\right>},
\end{align}
where we use the short hand
\begin{equation}
\left|\tilde F_J\right>=P\left|F_J\right>,
\end{equation}
for the particle-hole conjugate Slater determinant to $\left|F_J\right>$.

By setting $\left|X\right>=\left|F_J\right>$ and either $\left|Y\right>=\left|F_J\right>$ or
$\left|Y\right>=\left|\tilde F_J\right>$, we are left with the task of calculating
$\left<X\right|x_\alpha^\dagger x_{\beta'} H\left|Y\right>$. We focus on the term in $H$ containing
the density-density interaction between sites $-1$ and $0$. The remaining terms in the Hamiltonian
can be done using the same principles, but are simpler because they are single-particle additive
terms.

We note that $x_{\beta'}^\dagger x_\alpha\left|X\right>$ is a single Slater determinant. We can 
therefore apply (\ref{wick}) and obtain
\begin{eqnarray}
&&\left<X\right|x_\alpha^\dagger x_{\beta'} n_{-1} n_0\left|Y\right>\nonumber\\
&=&\frac{\left<X\right|x_\alpha^\dagger x_{\beta'} n_{-1}\left|Y\right>\left<X\right|x_\alpha^\dagger x_{\beta'} n_0\left|Y\right>}{\left<X\right|x_\alpha^\dagger x_{\beta'} \left|Y\right>}\nonumber\\
&&-\frac{\left<X\right|x_\alpha^\dagger x_{\beta'} c_{-1}^\dagger c_0\left|Y\right>\left<X\right|x_\alpha^\dagger x_{\beta'} c_0^\dagger c_{-1}\left|Y\right>}{\left<X\right|x_\alpha^\dagger x_{\beta'} \left|Y\right>}\label{intol}.
\end{eqnarray}

The expansion coefficients of $x_m$ in terms of the $c_m$ basis form a $(2\Omega+2)\times (2\Omega+2)$
real orthogonal matrix that we denote $X'$, such that
\begin{equation}
x_m^\dagger=\sum_{j=-1}^{2\Omega} c_j^\dagger X_{jm}'.
\end{equation}
The rectangular $(2\Omega+2)\times (\Omega+1)$ sub-block of $X'$ corresponding to its first $\Omega+1$ columns 
corresponds to the matrix $X$ defined in the previous appendix, i.e.
\begin{equation}
X_{j\alpha}=X_{j\alpha}'.
\end{equation}

Setting $\hat Z=x_\alpha^\dagger x_{\beta'}=\sum_{jk=-1}^{2\Omega}X_{j\alpha} X_{k\beta'}'
c_j^\dagger c_k$ and using (\ref{2ferm}) along with the fact that
$\sum_{j=-1}^{2\Omega} X_{j\nu} X_{j\alpha}=\delta_{\alpha\nu}$, we find that
\begin{equation}
\left<X\right|x_\alpha^\dagger x_{\beta'} \left|Y\right>
=\left<X\right|\left.Y\right>\left[(X')^\dagger Y M^{-1}\right]_{\beta'\alpha}.
\end{equation}
By further setting $\hat Z_2=c_j^\dagger c_k$ in (\ref{4ferm}) we find
\begin{align}
&\left<X\right|x_\alpha^\dagger x_{\beta'} c_j^\dagger c_k\left|Y\right>
=\frac{\left<X\right|x_\alpha^\dagger x_{\beta'} \left|Y\right>\left<X\right| c_j^\dagger c_k \left|Y\right>}
{\left<X\right|\left.Y\right>}
\nonumber\\
&+\left<X\right|\left.Y\right>\left[(X')^\dagger(1-YM^{-1}X^\dagger)\right]_{\beta'j}\left[YM^{-1}\right]_{k\alpha}.
\end{align}
When this is substituted back into (\ref{intol}) one obtains a formula for
$\left<X\right|x_\alpha^\dagger x_{\beta'} n_{-1} n_0\left|Y\right>$ that can be evaluated if $X'$
is known. (Both $Y$ and $M^{-1}$ can be calculated if the $X'$ of each term $J$ is known.) We obtain
$X'$ by diagonalizing $H_J$ numerically in the single-particle sector.

\section{Choice of parameters for the numerics}
\label{pars}
\begin{figure}[th]
\begin{center}
\includegraphics[width=.99 \columnwidth]{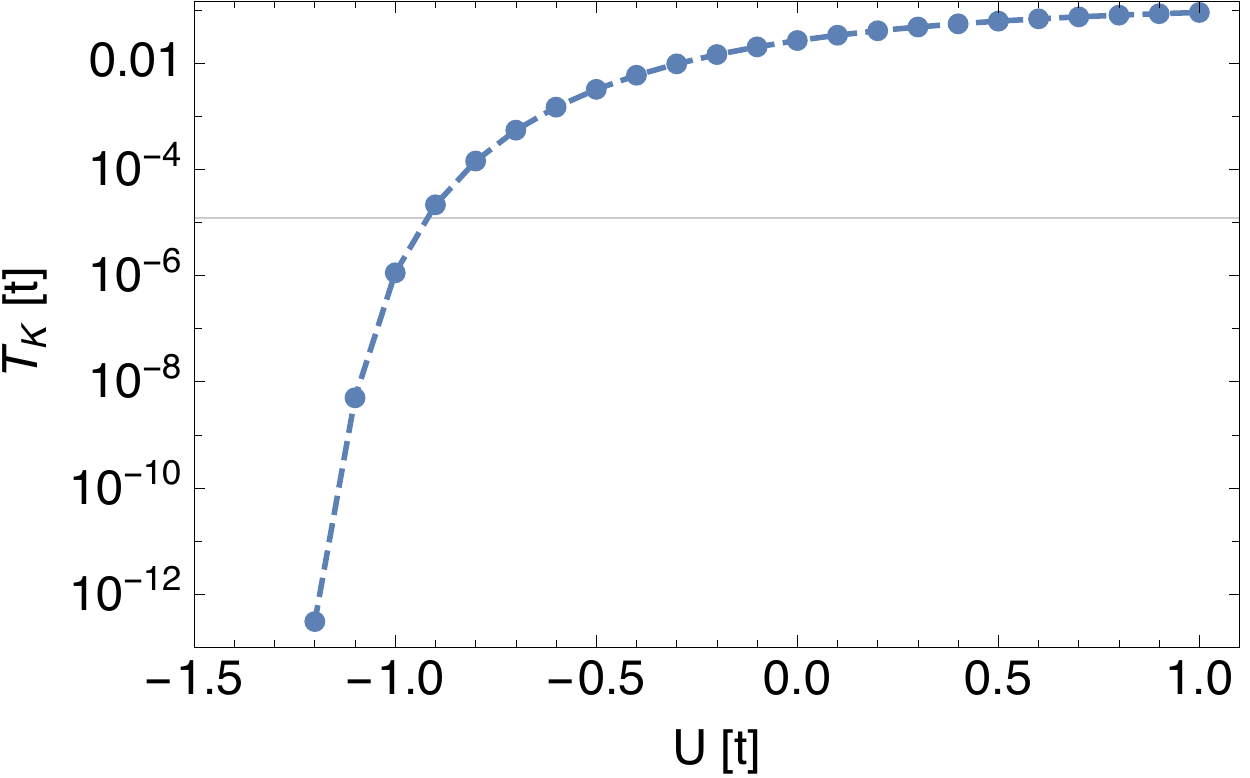}
\end{center}
\caption{Kondo temperature of the IRLM Wilson chain, versus $U$, calculated using NRG.
$\gamma=0.15\,t$, $\Lambda=1.5$\label{f1}}
\end{figure}

The following considerations informed our choice of the parameters $\gamma$ and $\Omega$. There
should be a reasonable separation of scales between the hybridization $\gamma$ and the band width
$2t$, otherwise universal many-body effects are obscured by non-universal ultraviolet effects. At
the same time, if $\gamma$ is too small, features in the energy landscape that are associated with
important many-body physics become very shallow. Our relatively unsophisticated minimization
procedure can easily miss such features. We find that $\gamma=0.15 t$ strikes a good compromise.
Because we want to be reasonably sure of coming close to the absolute minimum of the energy
landscape, we also avoid extremely long Wilson chains associated to exponentially small energy
scales. We nonetheless want to have a system that is large enough to host fully developed Kondo
correlations in a significant portion of the symmetric phase. We use $\Omega=28$, which translates
into a system with $29$ particles distributed among $58$ orbitals, and an infrared cut-off of
$10^{-5}t$. Exploring the $E_{\rm var}(\{f_J\},\{\varepsilon_n^{(J)}\},\{g_n^{(J)}\})$ landscape,
we find (not unexpectedly) that there are spurious local minima. However, we also observe that it is
not necessary to find the absolute minimum: there are many nearly degenerate minima that all give a
very reasonable approximation to the true ground state. In order to get a reasonable sampling of the
energy landscape, we do $125$ runs of a quasi-Newton (i.e. local) algorithm, with randomized
starting points, and take the overall lowest found minimum. This strategy works well for chains of
length up to $\sim60$ sites, and $M=2$ Slater pairs. If one wanted to access longer chains (lower
energies) or higher accuracy (more pairs), it seems one would need to adopt a more sophisticated
global minimization strategy. 

For $\Lambda=1.5$ and $\gamma=0.15 t$, we need to know at which value of $U$ the quantum phase
transition occurs. We also need to know at what value of $U$ in the symmetric phase, the emergent
energy scale $T_K$ becomes smaller than the infrared cut-off $10^{-5} t$. For this purpose, we
performed NRG on a very long chain, and calculated $T_K$ according to (\ref{tk}). Results are shown
in Fig. \ref{f1}. A horizontal line indicates the infrared cut-off associated with $\Omega=28$. We
see that the phase transition (where $T_K$ vanishes) occurs close to $U=-1.3 t$, and that $T_K$
equals the infrared scale at around $U=-0.9 t$. All further results presented in the main text are for
$\Omega=28$. We can expect to see fully developed strong correlations in the symmetric phase for 
$U \gtrsim -0.9 t$. For $-1.3 t<U<-0.9 t$, the correlations associated with Kondo physics are only 
partially developed.


%

\end{document}